\RequirePackage{amsmath}
\documentclass[runningheads]{llncs}
\usepackage{amsthm}
\usepackage[T1]{fontenc}
\usepackage{graphicx}
\usepackage{listings}
\usepackage{hyperref}
\usepackage{color}

\makeatletter
\renewcommand{\fnum@figure}{Listing \thefigure}
\makeatother

\makeatletter
\newcommand\notsotiny{\@setfontsize\notsotiny\@vipt\@viipt}
\makeatother

\graphicspath{{./problem1_images}}

\lstdefinelanguage[RISC-V]{Assembler}
{
  alsoletter={.}, % allow dots in keywords
  alsodigit={0x}, % hex numbers are numbers too!
  morekeywords=[1]{ % instructions
    lb, lh, lw, lbu, lhu, ld, li, la, 
    sb, sh, sw, sd,
    mv,
    sll, slli, srl, srli, sra, srai,
    add, addi, sub, lui, auipc,
    xor, xori, or, ori, and, andi,
    slt, slti, sltu, sltiu,
    beq, bne, blt, bge, bltu, bgeu,
    j, jr, jal, jalr, ret,
    scall, break, nop, ecall, call,
    fdiv.s, fcvt.s.lu, fcvt.lu.s,
    sext.w, slliw, 
  },
  morekeywords=[2]{ % sections of our code and other directives
    .align, .ascii, .asciiz, .byte, .data, .double, .extern,
    .float, .global, .half, .kdata, .ktext, .set, .space, .text, .word, .section, .size, .type, .option, .ident
  },
  morecomment=[l]{;},   % mark ; as line comment start
  morecomment=[l]{\#},  % as well as # (even though it is unconventional)
  morestring=[b]",      % mark " as string start/end
  morestring=[b]'       % also mark ' as string start/end
}

\lstdefinestyle{customc}{
  belowcaptionskip=1\baselineskip,
  breaklines=true,
  xleftmargin=\parindent,
  language=C,
  showstringspaces=false,
  frame=single, 
  basicstyle=\scriptsize\ttfamily, 
  numbers=left, 
  numbersep=5pt, 
  captionpos=b,
  keywordstyle==\scriptsize\ttfamily,
  commentstyle=\scriptsize\ttfamily,
}

\begin{document}
%-------------------------------------------------------------------------------

%don't want date printed
\date{}

\title{Software Mitigation of RISC-V Spectre Attacks}

% \author{}   % BLIND REVIEW
% \institute{}
\author{
Ruxandra Bălucea\inst{1}
\and
Paul Irofti\inst{1,2}
% {\rm Ruxandra Bălucea \qquad Paul Irofti} \\
% Research Center for Logic, Optimization and Security (LOS),
% Department of Computer Science, \\
% Faculty of Mathematics and Computer Science,
% University of Bucharest, Romania \\
% {\rm Emails: ruxandra.balucea@unibuc.ro, paul@irofti.net}
} % end author

\institute{LOS-CS-FMI, University of Bucharest, Romania \and
Institute for Logic and Data Science, Romania\\
\email{ruxandra.balucea@unibuc.ro}, \email{paul@irofti.net}}

\maketitle

%-------------------------------------------------------------------------------
\begin{abstract}
%-------------------------------------------------------------------------------
Speculative attacks are still an active threat today that,
even if initially focused on the x86 platform,
reach across all modern hardware architectures.
RISC-V is a newly proposed open instruction set architecture
that has seen traction from both the industry and academia in recent years.
In this paper we focus on the RISC-V cores
where speculation is enabled
and, as we show,
where Spectre attacks are as effective as on x86.
Even though RISC-V hardware mitigations were proposed in the past,
they have not yet passed the prototype phase.
Instead,
we propose low-overhead software mitigations
for Spectre-BTI,
inspired from those used on the x86 architecture,
and for Spectre-RSB,
to our knowledge the first such mitigation to be proposed.
We show that these mitigations work in practice
and that they can be integrated in the LLVM toolchain.
For transparency and reproducibility,
all our programs and data are made publicly available online.
\keywords{side-channel attacks \and hardware security \and system security}
\end{abstract}

%-------------------------------------------------------------------------------
\section{Introduction}

The introduction of Spectre~\cite{spectre} and Meltdown~\cite{meltdown} attacks in 2018
opened up a new field of research exploiting side-effects
that are spilled by speculation techniques inside the micro-architecture of modern processors~\cite{spectrev5,spectre_boom,kiriansky2018speculative,bhattacharyya2020specrop,ravichandran2022pacman,Wieczorkiewicz22_AMDBranch}.
Spectre attacks proved to be the hardest to mitigate~\cite{barberis2022branch,milburn2022you,Wieczorkiewicz22_AMDBranch},
even though it was attempted via both software~\cite{gruss2017kaslr,retpoline,windows_spectre,linux_spectre,nikolaev2022adelie}
and hardware~\cite{spectre_boom,koruyeh2020speccfi,martinoli2022cva6}
patches.
These attacks mainly targeted the popular x86 architecture,
but Spectre was later shown to affect multiple other architectures~\cite{spectre_boom,sabbaghsboomattacks,ravichandran2022pacman,miles2022simulating}.

RISC-V is a new open-standard instruction set architecture~(ISA)
\cite{riscvarch1}
recently proposed by University of California, Berkeley
that has seen wide academic and industry adoption 
\cite{martinoli2022cva6}.
In this paper we focus on reproducing and mitigating Spectre attacks on the RISC-V architecture.

%Having these prerequisites, we can understand the main problem of the speculation on BOOM.%
Even if the RISC-V cores are written from scratch
in order to research new efficient hardware methods,
they must also keep up with existing performance-inducing technologies.
Speculation is one of them
and it is present on all modern processors.
Despite recent speculation attacks,
unfortunately,
for mainstream architectures such as x86,
there are few hardware mitigations 
and even these
seem to not be sufficient~\cite{barberis2022branch}.
On RISC-V,
the few proposed hardware implementations ~\cite{spectre_boom,wistoff2022systematic,martinoli2022cva6} are mostly combinations or adaptations of the x86 ones. 
So, even if they seem to be quite efficiently in the present, as the RISC-V community grows, we expect the same problems as on x86.
In this context,
despite the fact that the same performance can not be achieved as with hardware solutions,
software mitigations remain the most practical and safe ones.

To our knowledge,
currently on RISC-V there are implemented the following variants of Spectre:
Spectre on Conditional Branches (Spectre v1),
Spectre Branch Target Injection (Spectre-BTI or Spectre v2) \cite{spectre_boom}
and
Spectre Return Stack Buffer (Spectre-RSB or Spectre v5)~\cite{sabbaghsboomattacks}.

In this paper we propose software mitigations for the Spectre-BTI variants and also for Spectre-RSB.
As far as we know,
this is the first time that Spectre-RSB mitigations are proposed.

Retpoline~\cite{retpoline} is such a mitigation for x86 that targets only Spectre-BTI.
As far as we know,
no software mitigation is known
for the RISC-V architecture and in fact, for any other RISC architecture.
We assume that this is also due to the fact that
for the RISC-V ISA
things are not as straight-forward as on x86
because the prologue and the epilogue of a function are more complex.
The stack frame requires saving of a really important callee-saved register - the return address \verb|ra|.
Retpoline is influenced by the calling-convention and how function return is achieved.
Therefore,
for RISC-V,
it can not be applied.
In this paper we propose a new software mitigation method for RISC-V that addresses and circumvents these issues.

Revisiting the main idea behind x86 Retpoline,
we note that 
this mitigation can be applied for Spectre v2
because speculation also appears in the context of a call instruction.
Thus,
we defend against this type of attack
by applying a defense technique derived from another speculation attack - Spectre v5.
The idea is that the indirect jump to an address from a register (x86 \verb|jmp|, RISC-V \verb|jalr|)
can be replaced with a direct call to a function (\verb|call|, \verb|jal|)
where the return address can be overwritten
with the value of that register.
At the return phase,
the execution will continue at the address from the register.
At the same time,
speculatively there will be executed the instructions under the call.
Thus,
in order to trap the speculation,
we add an infinite loop after
the indirect jump.
%\verb|jal|. 

Focusing on RISC-V,
this defense can not be applied in the same manner.
If we modify the return address with the desired register value,
the function called indirectly will also have as return address
the beginning of the function
and the execution will be caught in an infinite loop 
(we describe this in detail around Listings \ref{code:Retpoline-jump} and \ref{code:Retpoline-call}).
This is because the return is not dictated by the value from the top,
but by the return address register
which is saved on the stack and restored at the end
(we describe this behavior in detail around Listing~\ref{code:Prologue_Epilogue}).
Nevertheless,
this mitigation can be applied as described above
in specific contexts:
for indirect jumps
there is no stack frame created
and there is no dependency on the value of the return address register.

\paragraph{Contribution.}
Our main contribution is
the proposal of software mitigations on RISC-V against Spectre attacks.
To this end we provide an implementation of the proposed defense
that handles
Spectre-BTI, for both indirect jumps and calls,
and Spectre-RSB.
To our knowledge,
this is the first time that Spectre-RSB mitigation is proposed.
The distinction can be made directly in the assembly code
and the defense can be applied by
replacing the jump/call instructions with specific code.
To prove this,
we provide a publicly available LLVM feature
that can be activated at compilation time
through enabling the mitigations via a single flag.
The resulting executable can be run on the RISC-V speculative core BOOM.
Spectre-BTI and Spectre-RSB will be no longer reproduced.
Another contribution is the adaptation of
the existing Spectre variants for the RISC-V speculative cores
that we implement in practice and make publicly available.
We also provide the steps necessary to reproduce our research
together with our test programs and data.

\paragraph{Outline.}
In Section~\ref{sec:attacks},
we revisit and adapt the Spectre attacks
needed in order to prove that RISC-V is vulnerable to this type of attacks,
which are also required in part for our proposed mitigations.
Next,
in Section~\ref{sec:mitigations},
we introduce the proposed defenses against
Spectre-RSB and two types of Spectre-BTI attacks.
We test our attack and mitigations attacks and provide experiments
along with ways of reproducing our results in Section~\ref{sec:experiments}.
In the next section we conclude and make publicly available our implementation and data.

%-------------------------------------------------------------------------------
\section{Berkeley Out of Order Machine}
\label{sec:BOOM}
Berkeley Out of Order Machine (BOOM) \cite{boom,boomv2,boomv3} is an open-source RV64GC core written in Chisel. 
It is superscalar, out-of-order and speculative, being an ideal candidate for our work.
The speculation is dictated by a two-level branch predictor composed of a Next-Line Predictor (NLP) and a Backing Predictor (BPD). The predicted address is chosen based on two other structures incorporated in the NLP - Branch Target Buffer (BTB) and Return Address Stack (RAS). The taken/not taken decision is up to the BPD, but as we do not address an attack based on branches, we will not present more information here.

BTB is a table with $64 \times 4$ entries, set-associative which stores a mapping from a PC address to a target address. 
A tag search is initiated in this table,
whenever a prediction for an indirect jump is needed.

RAS is a stack which maintains in the top the following address after the last call. This value is popped when a \verb|ret| instruction is met. The stack structure was chosen in order to handle nested calls. However, this was a problem in the second version of BOOM because the stack was not updated correspondingly in case of a mispredict. This was solved in SonicBoom, the third version of BOOM.

%-------------------------------------------------------------------------------

\section{RISC-V Spectre Attacks}
\label{sec:attacks}

This section presents
Spectre-BTI (Branch Target Injection)~\cite{spectre}
and
Spectre-RSB (Return Stack Buffer)~\cite{spectrev5}
in the RISC-V context~\cite{spectre_boom}
along with the side-channel technique Evict\&Reload~\cite{evict_reload}
which is a prerequisite for these attacks. Both attacks are illustrated by reading memory from the same process, in-place, referred to as BTB-SA-IP and RSB-SA-IP accordingly to the threat model presented in \cite{canella2019systematic}. 
% Also, it is described the software mitigation for Spectre-BTI, presenting how X86 Google Retpoline can be applied for a totally different ISA with multiple indirect patterns.

%-------------------------------------------------------------------------------
\subsection{Spectre-BTI}
Spectre-BTI was reproduced on RISC-V on the experimental speculative core BOOM.
In this variant, arbitrary locations in the allocated memory of a program can be read exploiting the indirect branch instructions - \verb|jalr| for calls and \verb|jr| for jumps. Each jump/call to an indirect address, loaded in a register, creates a speculation window during which essential information can be brought into the cache memory.
As on other architectures, in case of a mispredict, the cache is not cleared and the information can be retrieved by an attacker.

The attack is illustrated by reading memory from the same process,
having a role-play between an attacker and a victim.
In our experiments
we use this approach due to the limitations imposed by the simulator (as will be later described). The time needed to execute is quite long,
so we prefer to use a single binary.
In the first phase,
the attacker mistrains the Branch Target Buffer (BTB)
jumping for a large number of times to a valid fixed address.
The valid jump is taken to a segment of code that
discloses information from a certain memory region.
This step makes the predictor assume that the jump will always be taken.
In the second stage,
the attacker makes the victim execute an indirect jump to another (normally illegal) address,
where the disclosed information is of interest to the attacker,
and,
due to the training phase and speculation,
the predictor assumes the jump will be taken
and the pipeline proceeds with the memory access.
Thus,
the second phase can create side-effects into the cache,
side-effects that provide unauthorized information  to the attacker.
In the end, even if the jump is made to the correct address, the data from cache can still be read by the attacker.

We will present here only the main aspects of this attack
in order to introduce our work.
The implementation details can be found in the Supplementary Material
and also in the original paper \cite{spectre_boom}.
Spectre authors present an attack based on
the indirect calls having two pieces of code similar to the functions presented in Listing \ref{code:Spectre-v2}.
Spectre-v2 was presented by the authors only for indirect calls
that appear, for example,
when we are talking about virtual functions.
We extended this example and add a new one for the indirect jumps when the register keeps the address of a snippet of code, such as for a switch case.
Thus, in the new example,
we took the assembly code generated for this function,
removed the instructions related to the stack frame
and used the global variable \verb|passInIdx|
to access the desired memory. Even if for the calls we could have maintained \verb|passInIdx| as a parameter, we also kept it as a global variable for linearity.

%-------------------------------------------------
\begin{figure}[t]
\begin{lstlisting}[xleftmargin=3mm]
uint64_t passInIdx;
uint8_t array1[10] = {1,2,3,4,5,6,7,8,9,10};
uint8_t array2[256 * L1_BLOCK_SZ_BYTES];
char* secretString = "BOOM!";

void wantFunc(){
    asm("nop");
}

void victimFunc(){
    temp &= array2[array1[passInIdx] * L1_BLOCK_SZ_BYTES];
}

int main() {
    uint64_t attackIdx = 
       (uint64_t)(secretString - (char*)array1);
     ...
     //victimFunc address is loaded in %[addr] 
     //    for the training phase
     // wantFunc addrees is loaded in %[addr] 
     //    by the victim
     "jalr ra, \%[addr], 0\n"
     ...
}
\end{lstlisting}
\caption{Spectre v2}
\label{code:Spectre-v2}
\end{figure}
%-------------------------------------------------

As presented above,
the BTB is trained in the first stage to predict the \verb|victimFunc| address.
The jump to that function was repeated 40 times,
each time assigning different valid values to the \verb|passInIdx| variable.
The 41st time,
as it can be seen in line 15,
the attacker assigned to this variable a convenient value,
for example, the index corresponding to the beginning of the secret.
In the second phase, in line 22,
the victim tries to call via an indirect instruction \verb|wantFunc|,
but speculatively  \verb|victimFunc| is called again.
So, in line 11, \verb|array2[array1[attackIdx] * L1_BLOCK_SZ_BYTES]| is  brought in the cache
(i.e. \verb|array2['B' * L1_BLOCK_SZ_BYTES]|).
Having this value in the cache and access to \verb|array2|,
the attacker can retrieve the first character from the password
with a side-channel attack method such as Evict \& Reload~\cite{evict_reload}.
For your convenience, we review this in the Supplementary Material.

For more details,
the reader is advised to consult the full attack
provided in the Supplementary Material. 
There,
the code presented in Listing \ref{code:full-spectre-attack}
is for an attack on indirect calls
(see the called functions from Listing~\ref{code:functions}). 
For indirect jumps,
at line 73,
we should have a jump instruction: \verb|jalr x0, %addr, 0|. 
Also,
for the return from the snippets of code presented in the assembly file from Listing \ref{code:gadgets},
we added at the end a jump back to a label from the source file. 
This label should be added after the indirect jump at line 74
and declared as global before main (\verb|asm(".global end\n")|).

\subsection{Spectre-RSB}
Spectre-RSB \cite{spectrev5},
known as Spectre-v5,
was reproduced on SonicBoom,
the third generation of BOOM
which added as a feature a functional RAS.
In this variant,
the vulnerability is based on the RAS hardware stack
where the most probable return addresses are pushed
for each \verb|call| instruction.
Based on these values,
the return from a function is speculatively computed and,
as before,
a speculation execution window is created.
Although,
if the value of the return address register \verb|ra|
is manipulated during the function,
the program will continue the execution on a different path
and the information brought into the cache by the instructions executed speculatively will not be erased.
In this context,
again,
an attacker can retrieve the information using the Flush \& Reload technique.

%-------------------------------------------------------
\begin{figure}[t]
\begin{lstlisting}[xleftmargin=3mm]
__asm__ (
    "frameDump:";
    "# Pop off stack frame and get main RA"
    "ld ra, 56(sp)";
    "addi sp, sp, 64";
    "ld fp, -16(sp)";
    ...
    "ret");
void specFunc(char *addr){
    extern void frameDump();
    uint64_t dummy = 0;
    frameDump();
    char secret = *addr;
    dummy = array2[secret * L1_BLOCK_SZ_BYTES]; 
    dummy = rdcycle();
}
\end{lstlisting}
    \caption{Spectre v5}
    \label{code:Spectre-v5}
\end{figure}
%-------------------------------------------------------

For BOOM,
the implementation of RAS generates a new stack entry: the address of the next instruction after the \verb|call|.
In Listing~\ref{code:Spectre-v5} we illustrate the attack.
As can be seen,
it is enough to add a function which modifies the return address
and add relevant code after the call to this function (lines 13-15).
To accomplish this,
the function \verb|frameDump| (line 2)
loads in \verb|ra| the value of the return address of the function \verb|specFunc| (line 4)
and the stack frame is popped (line 5),
so the execution will continue directly in the calling function of \verb|specFunc|.

Similar to what we discussed in the previous attack,
the attacker can set the parameter to \verb|specFunc| as the desired address (line 9),
in this case the address of the secret string.
The value from \verb|array2| (line 14)
corresponding to the first character
will be brought into memory
and the attacker will be able to retrieve the information
using Flush \& Reload.
By repeating the attack for all characters,
the secret will be revealed.

\section{RISC-V Spectre Mitigations}
\label{sec:mitigations}

Given the attacks from Section~\ref{sec:attacks},
we now propose 
two Spectre-BTI mitigation strategies for the RISC-V architecture,
inspired by the x86-specific software mitigation Retpoline~\cite{retpoline}
and a new Spectre-RSB mitigation,
the first in the field as far as we know.
In the current section
we present and discuss ways of replacing
indirect jumps and calls
with a sequence of instructions that
will provide the same behavior
while removing the speculation attack.

\subsection{Spectre-BTI: Indirect Jumps}
Indirect jumps are realized using the \verb|jr| instruction
which is in fact an assembly pseudo instruction for \verb|jalr|
with the first operand set as register \verb|X0|.
$$
\text{jr rd, rs1} \rightarrow \text{jalr x0, rs1, 0}
$$
This register is hardwired zero.
So, its presence on that position indicates that no register will take the value of the following instruction address.

The mitigation is summarized in Listing~\ref{code:Retpoline-jump};
the first block represents the original indirect jump
and the second its replacement.
To replace the \verb|jr| instruction (first block, line 1),
we use the Spectre v5 vulnerability
and rewrite it as a direct call to a pseudo-function
with no calling-convention applied (second block, line 1).
In this function we store in \verb|ra| the value of the register from the indirect jump (line 5).
At the end we do a \verb|ret| - an indirect jump to the return address register \verb|jr ra| (line 6).
During this time the speculation will be caught in an infinite loop
that takes place after the \verb|call| instruction
(lines 2--3).
\begin{remark}
Regarding line 6, it may seem that the original problem from line 1 was only moved below due to the usage of the same instruction (the unconditional jump \verb|jr|). In fact this is not the case because this new jump has a special property - it is a return instruction. The unconditional jumps having as operand the register \verb|ra| are marked as \verb|ret|s and are used only to remove the RAS entry added by the calls. It would make no sense to predict a target of a \verb|ret| as it depends on the location of the associated call.
This behavior was also confirmed by our experiments from Section 5.
\end{remark}
%--------------------------------------
\begin{figure}[t]
\begin{minipage}{0.3\columnwidth}
\setcounter{figure}{3}
\renewcommand{\thelstlisting}{\arabic{figure}}
\begin{lstlisting}[language={[RISC-V]Assembler},xleftmargin=3mm]
jr	a5
\end{lstlisting}
\begin{lstlisting}[language={[RISC-V]Assembler},xleftmargin=3mm,caption={RISC-V mitigation - indirect jump},captionpos=b, label={code:Retpoline-jump}]
jal set_up_target
capture_spec:
  j capture_spec
set_up_target:
  addi ra, a5, 0
  jr ra
\end{lstlisting}
\end{minipage}
%\caption{RISC-V mitigation - indirect jump}
\hfill
\begin{minipage}{0.65\columnwidth}
\setcounter{figure}{4}
\renewcommand{\thelstlisting}{\arabic{figure}}
\begin{lstlisting}[language={[RISC-V]Assembler},xleftmargin=3mm]
addi sp, sp, -16 # add space on the stack
sd ra, 8(sp)     # save the return address
sd fp, 0(sp)     # save the frame pointer
addi fp, sp, 16  # modify the stack frame base
\end{lstlisting}
\begin{lstlisting}[language={[RISC-V]Assembler},xleftmargin=3mm,caption={Current general function prologue (top) and epilogue (bottom)},captionpos=b, label={code:Prologue_Epilogue}]
ld fp, 0(sp)     # restore the frame pointer
ld ra, 8(sp)     # restore the return address
addi sp, sp, 16  # reduce the size of the stack
jr ra            # return in the caller
\end{lstlisting}
\end{minipage}
%\caption{Current general function prologue (top) and epilogue (bottom)}
\end{figure}
%--------------------------------------

\subsection{Spectre-BTI: Indirect Calls}

For the indirect calls, the transformation is not so simple.
The indirect calls are reflected in the \verb|jalr| single-operand pseudo-instruction
which is an alias for the instruction with the same name, but more operands.
$$
\text{jalr rs1} \rightarrow \text{jalr ra, rs1, 0}
$$
The first operand
which is the operand that will take the value of the following instruction address
is in this case set by default to \verb|ra|.
In this way,
the return from the called function is right after the call instruction
and now it is quite clear why this value is chosen as a RAS entry.
\begin{align*}
\text{ra} &\leftarrow \text{pc} + 4 \\
\text{pc} &\leftarrow \text{rs1} + 0    
\end{align*}
In order to achieve the same behavior as for the indirect jumps
we need to find a way not to overwrite the return address
for the functions called through the register.
We want to maintain the idea of
overwriting the return address for the \verb|set_up_target| function
with the address of the beginning of the function stored in the register.
Thinking about where does the called function return,
we discover that in fact that address is not represented by the value from \verb|ra|,
but by the value from the stack restored at the end in \verb|ra|.
Thus we can replace the return address register
with the value of the register from the indirect call,
but with one condition:
we can not store this new address on the stack. Instead, we need to save the legitimate one - the address after the indirect call.

\begin{remark}
If during the function execution the return address register \verb|ra| is modified, for example when handling an error via an early return inside an if-clause, our mitigation will not affect the normal program behavior.
\end{remark}

In Listing \ref{code:Prologue_Epilogue}
we present an usual prologue and epilogue
for a 64-bit RISC-V core.
In the Prologue (top block),
in order to meet the condition presented above,
we need to jump
over the instruction that adds space on the stack by default (line 1)
and
over the instruction that stores the value of \verb|ra| on the stack (line 2).
In order to do this,
we need to recreate these instructions in the body of the \verb|set_up_target|.

In practice the first lines in the prologue
are not always the ones presented in the top block of 
Listing~\ref{code:Prologue_Epilogue}.
These lines are changed by adding the callee-saved registers on the stack.
These are resizing the stack and the space added becomes dependent on their number.
For example,
for a given function \verb|f1|,
registers \verb|s1| and \verb|s2| must be saved on the stack
so the allocated space is increased to 32 bytes.
Another function \verb|f2|,
that is also called indirectly,
requires a single register to be saved and the allocated space is only of 24 bytes.
Our goal is to replace the indirect call with the same code all the time
no matter of the function at hand.

%--------------------------------------
\begin{figure}[t]
\begin{minipage}{0.46\textwidth}
\begin{lstlisting}[language={[RISC-V]Assembler},xleftmargin=3mm]
addi sp, sp, -32
sd ra, 24(sp)       
sd fp, 16(sp)        
addi fp, sp, 16  
sd s1, 8(sp)
sd s2, 0(sp)
\end{lstlisting}
\end{minipage}\hfill
\begin{minipage}{0.46\textwidth}
\begin{lstlisting}[language={[RISC-V]Assembler},xleftmargin=3mm]
<@\textcolor{red}{addi sp, sp, -16} @>    
<@\textcolor{red}{sd ra, 8(sp)} @>
<@\textcolor{red}{sd fp, 0(sp)} @>      
<@\textcolor{red}{addi fp, sp, 0} @>    
<@\textcolor{red}{addi sp, sp, -16} @>  
sd s1, 8(sp)
sd s2, 0(sp)
\end{lstlisting}
\end{minipage}
\caption{Prologue mitigation for function \texttt{f1}:
top block represents the original prologue
and the bottom block presents the proposed mitigation.}
\label{code:Prolog change}
\end{figure}
%--------------------------------------

Thus the first measure to be taken is one that offers consistency to the instructions used by the prologue.
We propose to accomplish this in two separate phases.
The idea here is to modify the prologue of all functions
such that in the first phase,
the memory is allocated only for the registers saved all the time - \verb|ra| and \verb|fp|.
In the second stage,
the stack size can be adjusted by the initial value minus 16 bytes
(in case of a 64-bit architecture).
From then on,
the compiler can continue to emit the stores for the other callee-saved
and the rest of the function body.
Therefore,
the initial part of the prologue is replaced
by one with the same behavior which keeps the first instructions constant.

As an example,
the transformation for the \verb|f1| function
is presented in Listing \ref{code:Prolog change}. In the first frame, the stack allocation is the usual one, similar to the one exposed in Listing \ref{code:Prologue_Epilogue}, adapted for the \verb|f1| function. In the second frame, the prologue is changed as previously described. The stack size is initially increased only by 16 bytes (line 1) in order to allocate space for the storage of \verb|ra| and \verb|fp| (lines 2 - 3). Now, the frame pointer is modified to point to the value of the old \verb|fp| by taking the value of \verb|sp| (line 4). As a last step, at line 5, the value of \verb|sp| is decreased again with the necessary amount of space for the callee-registers - 16 bytes for \verb|s1| and \verb|s2| (the stack grows downwards).   
%--------------------------------------
\begin{figure}[t]
\begin{minipage}{0.46\textwidth}
\setcounter{figure}{6}
\renewcommand{\thelstlisting}{\arabic{figure}}
\begin{lstlisting}[language={[RISC-V]Assembler}, xleftmargin=3mm]
jalr	a5
\end{lstlisting}
\begin{lstlisting}[language={[RISC-V]Assembler}, xleftmargin=3mm,caption={RISC-V mitigation - indirect call},captionpos=b, label={code:Retpoline-call}]
jal set_up_target
capture_spec:
  j capture_spec
set_up_target:
  addi ra, a5, 4
  addi sp, sp, -16
  la a5, end
  sd a5, 8(sp)
  jr ra
end:
\end{lstlisting}
\end{minipage}
\hfill
\begin{minipage}{0.46\textwidth}
\setcounter{figure}{7}
\renewcommand{\thelstlisting}{\arabic{figure}}
\begin{lstlisting}[language={[RISC-V]Assembler}, xleftmargin=3mm]
call	frameDump
\end{lstlisting}
\begin{lstlisting}[language={[RISC-V]Assembler}, xleftmargin=3mm, caption={RISC-V mitigation - Spectre RSB},captionpos=b, label=code:Retpoline-RSB]
jal set_up_target
capture_spec:
  j capture_spec
set_up_target:
  la ra, frameDump
  jr ra
\end{lstlisting}
\end{minipage}
\end{figure}
%--------------------------------------
We generalize this approach
and
introduce the resulting instructions in the body of
the \verb|set_up_target| function.
The full implementation is depicted in Listing~\ref{code:Retpoline-call}:
the top block contains the original indirect call instruction
and the bottom block our proposed mitigation.
%Now that we have these changes applied, the instructions can be reproduced in the body of the \verb|set_up_target| function.
On line 5,
in order to jump over the first two instructions,
we need to add in \verb|ra| the value from the register plus 4.
For this, we remind the reader that
we use RV64GC - the default target for the existing compilers.
In this case,
some instructions
like \verb|addi| and \verb|sd|
are compressed on 2 bytes each.
After that,
on line 6,
we need to add the instruction which allocates space
for the registers \verb|ra| and \verb|fp|
%(line 4, Listing \ref{code:Retpoline-call})
and
store on the stack (lines 7--8) the address at the end of the snippet of code
(line 10).
In our LLVM implementation
we computed the offset for the relative jump,
but here,
for clarity,
we store the address of a pre-added label (line 10).
Other than that,
the idea is the same as for the indirect jump,
the call to the function is realized using the value from the \verb|ra| register
(line 9)
and the speculation is trapped after the call (lines 2--3).

\begin{remark}
The transformation presented in \ref{code:Retpoline-call} is applied in case of using the compressed extension. Also, the function and the call should be in files compiled with the same option (with or without the compressed extension activated). 
\end{remark}

% %https://download.vusec.net/papers/bhi-spectre-bhb_sec22.pdf - We
% conclude software defenses such as retpoline remain the only
% practical BTI mitigations in the foreseeable future and the
% pursuit for efficient hardware mitigations must continue.

\subsection{Spectre-RSB}

The idea behind this mitigation is similar to the one presented for the two variants of Spectre-BTI. 
We need to avoid a \verb|call| instruction which will add into the RAS an address that will be used for speculation.

A call does not have as an operand a register, but a relocated symbol whose address is either known, either will be computed at link time. 
Either way, there is no reason not to use the symbol in a different instruction. 
So, 
similar to moving the value of the register used for indirect jumps in \verb|ra|, we can use the symbol for a load in \verb|ra|.

As a result,
we propose a mitigation where,
as per Listing \ref{code:Retpoline-RSB},
we maintain the idea of catching the speculation in an infinite loop (lines 
2 - 3) and make a call to the \verb|set_up_target| function (line 1).
In this function
with no prologue and no epilogue,
we load the address of the symbol in the \verb|ra| register(line 5)
and return basically at the beginning of the function that we need to call (line 6).

%-------------------------------------------------------------------------------
\section{Experiments\protect\footnotemark}
\label{sec:experiments}

\footnotetext{Programs, code and data
available at \url{https://github.com/riscv-spectre-mitigations}
}
To run our experiments we used a superscalar, speculative, out-of-order core named BOOM (Berkeley Out-of-Order Machine).  For this project we used the latest version of BOOM named SonicBoom. BOOM can be also integrated in a SoC using the majority of hardware structures from Rocket Chip by loading them like a library.
BOOM can be used as a part of a larger project named Chipyard which includes a number of different cores, tools, accelerators and simulators. From this project, different configurations of a chip can be generated with different numbers of cores, with vectorization support or different number of inputs for certain components. In our experiments, we used the smallest available configuration - \verb|SmallBoomConfig|. 

These configurations can be used directly on FPGAs or using the VCS simulator.
They can also be executed on the open-source simulator Verilator which was our choice as well.
Being a software simulated environment,
execution times can take a really long time.
Nevertheless, the results are reliable and the behavior is similar as for the other options.
Even though we reached out to other vendors that offer RISC-V chips with speculation enabled,
in our case this was the only testbed available
that we could run our attacks
and test our proposed mitigations on.
To reproduce our experiments,
we created a minimal configuration
in the \verb|Spectre-v2-v5-mitigation-RISCV| repository.
The interested reader
should also consult the official documentation
of BOOM \cite{boomv3} and Chipyard \cite{chipyard}.

The mitigations for the scenarios presented in Section~\ref{sec:mitigations}
were adapted and integrated in the LLVM toolchain.
In the future,
we hope to get our work integrated in the official LLVM project.
The patchset and the full tree of the modified LLVM version
is also made available online in our repository.
To reproduce our results,
it is necessary to download the updated version of LLVM%
and build it following the recommendations on their official page.
Additionally,
GNU toolchain version 2.32 for RISC-V
%\footnote{\url{https://github.com/riscv-collab/riscv-gnu-toolchain}}
needs to be installed in the same directory as LLVM.

Our repository also contains programs
testing for and, if possible, reproducing the attacks 
for the two variants of Spectre v2,
on indirect jumps (see \verb|indirectBranchSwitch|),
and indirect calls (see \verb|indirectBranchFunction|) and also for Spectre v5 (\verb|returnStackBuffer|).
These can be compiled and executed using the Makefile.
To activate the mitigation
it is necessary to add the parameter \verb|RETPOLINE=1|
to the make command.
For both cases,
there are also some variants of the tests
that do not need the updated compiler.
Here,
the attack is mitigated directly from the code,
using inline assembly and manually replacing
the unsafe sections as described in Section~\ref{sec:mitigations}.

%--------------------------------------------------
\lstset{
 basicstyle=\notsotiny\ttfamily}
\begin{figure*}[t]
\begin{minipage}{0.48\textwidth}
\begin{lstlisting}[language={bash}, xleftmargin=3mm]
./simulator-chipyard-SmallBoomConfig bin/indirectBranchFunction.riscv
The attacker guessed character B 8 <@times@>.
The attacker guessed character O 8 <@times@>.
The attacker guessed character O 7 <@times@>.
The attacker guessed character M 8 <@times@>.
The attacker guessed character ! 9 <@times@>
The guessed secret is BOOM!
./simulator-chipyard-SmallBoomConfig bin/indirectBranchSwitch.riscv
The attacker guessed character B 7 <@times@>
The attacker guessed character O 6 <@times@>
The attacker guessed character O 7 <@times@>
The attacker guessed character M 6 <@times@>.
The attacker guessed character ! 8 <@times@>
The guessed secret is BOOM!
./simulator-chipyard-SmallBoomConfig bin/returnStackBuffer.riscv
The attacker guessed character B 9 <@times@>
The attacker guessed character O 8 <@times@>
The attacker guessed character O 6 <@times@>
The attacker guessed character M 6 <@times@>.
The attacker guessed character ! 10 <@times@>
The guessed secret is BOOM!
\end{lstlisting}
\end{minipage}\hfill
\begin{minipage}{0.48\textwidth}
\begin{lstlisting}[language={bash}, xleftmargin=3mm]
./simulator-chipyard-SmallBoomConfig bin/indirectBranchFunction.riscv
The attacker guessed character  1 <@times@>.
The attacker guessed character  1 <@times@>.
The attacker guessed character  1 <@times@>.
The attacker guessed character  1 <@times@>.
The attacker guessed character  1 <@times@>.
The guessed secret is 
./simulator-chipyard-SmallBoomConfig bin/indirectBranchSwitch.riscv
The attacker guessed character  1 <@times@>.
The attacker guessed character  1 <@times@>.
The attacker guessed character  1 <@times@>.
The attacker guessed character  1 <@times@>.
The attacker guessed character  1 <@times@>.
The guessed secret is 
./simulator-chipyard-SmallBoomConfig bin/returnStackBuffer.riscv
The attacker guessed character  0 times.
The attacker guessed character  1 times.
The attacker guessed character  0 times.
The attacker guessed character  1 times.
The attacker guessed character  0 times.
The guessed secret is 

\end{lstlisting}
\end{minipage}
\caption{Attacks (left) and mitigations (right):
spectre attack is repeated 10 times for each memory read.
Left block recovers the seceret "BOOM!" via three Spectre attacks;
right block attempts to do the same but with mitiagtions enabled but fails.}
\label{code:results}
\end{figure*}
%--------------------------------------------------
We present an instance of our experiments in Listing~\ref{code:results}
where the left block reproduces the Spectre attacks
and the right block tries to reproduce them with
mitigations enabled thus failing to retrieve the secret.
As customary with Spectre attacks,
due to the empirically chosen cache hit threshold,
the confidence level of the retrieved data is increased by running the attack for ten times
on each character from the secret.
As we can see in Listing \ref{code:results}
in the left block,
on an unpatched system,
the characters are guessed in the majority of times.
After adding the LLVM compiler option that includes our mitigations,
in the right block of Listing \ref{code:results},
the characters are no longer guessed.
Nothing will be printed in the console,
as each time
a different non-printable character from the ASCII code is guessed.
Other times no character is guessed at all
(denoted "0 times" in the figure)
as nothing was found in the cache.
This is why we do not see a character in the output
and this is also why for each character
we get that it was guessed only a single time.

Regarding the performance impact of our proposed mitigations,
unfortunately,
using the simulator as our only option,
did not permit us to obtain a reliable execution time performance analysis.
Of course,
the code size will be increased by the instructions
depicted in Listings~\ref{code:Retpoline-jump} and \ref{code:Retpoline-call},
but we argue that this small increase is acceptable.

The code size depends on the usage of the compressed extension (RV64GC). 
Also, 
the size difference is influenced by the number of
indirect jumps,
indirect calls,
direct calls,
and functions. 
The number of bytes for each case is presented in Table \ref{tab:size_dif}. 
For indirect jumps and calls,
the difference results from adding extra instructions as presented in Listings \ref{code:Retpoline-jump} and \ref{code:Retpoline-call}. 
For functions,
only one supplementary instruction is added
by splitting the stack allocation in two phases.
%------------------------------
\begin{table}[t]
    \centering
    \begin{tabular}{c|rr}
      %\hline
       &RV64G & RV64GC \\
      \hline
      Indirect jumps &  12 bytes & 10 bytes\\
      %\hline
      Indirect calls & 28 bytes & 22 bytes\\
      %\hline
      Function Prologue & 4 bytes & 2 bytes \\
      %\hline
       Direct calls & 16 bytes & 14 bytes \\
      %\hline
    \end{tabular}
    \caption{Size difference for each change created by the mitigation for the standard ISA (RV64G) and standard ISA with the  compressed extension (RV64GC).}
    \label{tab:size_dif}
\end{table}
%------------------------------
Future research can help reduce this code size increase by employing
static or dynamic analysis to identify and replace only the vulnerable paths.
Given that our mitigations have a similar approach to
that of the x86 Retpoline implementation
which is in use by most users today,
we expect this to also be the next step for RISC-V development
and to become the default on this platform.
Nowadays kernels on x86 are compiled with this mitigation for 
both Windows \cite{windows_spectre}
and Linux (since 4.15) \cite{linux_spectre}
operating systems.
Also,
the Retpoline authors showed that this mitigation 
does not cause significant performance degradation for x86 \cite{Performance_Google}.

%-------------------------------------------------------------------------------
\section{Conclusions}
In this paper
we reproduced Spectre-BTI and Spectre-RSB attacks
on the RISC-V speculative core BOOM.
Our main contribution represents the proposed software mitigations
for Spectre-RSB,
to our knowledge the first mitigation for this attack,
and
for Spectre-BTI indirect jumps and indirect calls.
% that we demonstrate to be effective against
We demonstrate that these mitigations are effective against
Spectre variants
as depicted by our experiments.
The resulting work is integrated in the LLVM toolchain
for ease of use and reproducibility.

% %-------------------------------------------------------------------------------
% \section*{Acknowledgments}
% %-------------------------------------------------------------------------------
% The authors would like to thank the RISC-V Foundation for their kind hardware donation.

\clearpage
%-------------------------------------------------------------------------------
\bibliographystyle{plain}
\bibliography{bib}

%-------------------------------------------------------------------------------
\clearpage

\renewcommand{\thesection}{A}
\section{Supplementary Material}
\label{sec:supplementary-material}
\setcounter{subsection}{0}
\renewcommand{\thesubsection}{A.\arabic{subsection}}
\subsection{Evict \& Reload}
Evict \& Reload \cite{evict_reload} is a side-channel attack used to monitor the access to shared memory by timing the cache hits. 
The attacker can flush specific lines from cache and wait for a victim access.
After this event occurs,
the attacker can reload the memory lines measuring the time to load.
If the elapsed time is short,
it means that the the victim has already accessed that line and the information is stored into the cache.
Otherwise,
it will take longer because the line has to be brought from the main memory.

To flush a line from the cache memory,
the x86 ISA defines a special instruction \verb|clflush|.
%- SDM ???.
For RISC-V there is no such instruction
and the authors of \cite{spectre_boom}
had to implement a function with similar behavior.
The main difference between the two
is that this function evicts an entire set from the cache
and a set contains more than one line.
Thus, in order to reproduce the attack,
the shared memory must store elements at indexes multiple of the size of a set \verb|L1_ BLOCK_SZ_ BYTES|.
In addition,
the BOOM replacement policy is 4-way associative,
meaning that a memory block can occupy any of the 4 cache lines.
This means that in our function
the set must be flushed by \verb|4 * L1_WAYS|
where \verb|L1_WAYS| is the number of ways.
This value will assure that the set is indeed evicted.
The authors mention that by choosing this number,
the probability of eviction is \verb|99%|~\cite{spectre_boom}. 

In our case,
for the Spectre-v2 attack,
in the training phase,
the attacker will also flush \verb|array2| from the cache memory.
Now,
going back to the loading that occurs speculatively in \verb|victimFunc|,
we can use the reload step (see Supplementary Material, Listing~\ref{code:full-spectre-attack}, lines 78-83).
With this,
the attacker can find out the element that was accessed by the victim from \verb|array2|.
By accessing all the values from 0 to 256
(the ASCII codes for all the characters)
multiplied by \verb|L1_BLOCK_SZ_BYTES|,
the attacker can discover the index of the element accessed by the victim.
The time taken to load \verb|array2[66 *L1_BLOCK_SZ_BYTES]| (Listing \ref{code:Spectre-v2}, line 11)
will be much shorter because 66 is the ASCII code for the character \verb|'B'|,
which is the value used by the victim as well.
Therefore,
the attacker will discover the first character from the secret.

\newpage

\subsection{Full Spectre Attack}
\setcounter{figure}{9}
\renewcommand{\thelstlisting}{\arabic{figure}}
\begin{lstinputlisting}[label={code:full-spectre-attack},
caption={RISC-V Full Spectre Attack adapted from~\cite{spectre}.},
language={C}, xleftmargin=3mm]
{sources/indirectBranchFunction.c}
\end{lstinputlisting}

\newpage

\setcounter{figure}{10}
\renewcommand{\thelstlisting}{\arabic{figure}}
\begin{lstinputlisting}[label={code:functions},
caption={Extern functions used for the indirect calls.}
,
language={[RISC-V]Assembler}, xleftmargin=3mm]{sources/functions.s}
\end{lstinputlisting}

\setcounter{figure}{11}
\renewcommand{\thelstlisting}{\arabic{figure}}
\begin{lstinputlisting}[label={code:gadgets},
caption={Extern snippets of code used for the indirect jumps.}
,
language={[RISC-V]Assembler}, xleftmargin=3mm]{sources/gadgets.s}
\end{lstinputlisting}
%--------------------------------------

%-------------------------------------------------------------------------------
\end{document}